# An Exploratory Study on the Implementation and Adoption of ERP Solutions for Businesses

Jitesh Kumar Arora, Emre Erturk
Eastern Institute of Technology, New Zealand

## Abstract

Enterprise Resource Planning (ERP) systems have been covered in both mainstream Information Technology (IT) periodicals, and in academic literature, as a result of extensive adoption by organisations in the last two decades. Some of the past studies have reported operational efficiency and other gains, while other studies have pointed out the challenges. ERP systems continue to evolve, moving into the cloud hosted sphere, and being implemented by relatively smaller and regional companies. This project has carried out an exploratory study into the use of ERP systems, within Hawke's Bay New Zealand. ERP systems make up a major investment and undertaking by those companies. Therefore, research and lessons learned in this area are very important. In addition to a significant initial literature review, this project has conducted a survey on the local users' experience with Microsoft Dynamics NAV (a popular ERP brand). As a result, this study will contribute new and relevant information to the literature on business information systems and to ERP systems, in particular.

## Project Aims

ERP systems can provide support for all functions of an organisation and can help conduct business operations more efficiently, through their large online databases. ERP systems are used to track business assets (for example, raw materials, capital assets, and money) and to monitor procurement requests and financial transactions. As a result, they can also reduce production times and costs. These systems handle the data input for many business processes. ERP systems have seen rapid adoption in all of the major industries, for example, public insurance, assembly, retail, and education. Businesses in New Zealand also need to manage the productivity of their core processes. For example, in schools and universities, ERP systems and databases cover finance, accounting, human resources, logistics, enrolments, admissions, and so on. Besides, the need to centralise and secure organisational data is an important concern that may lead to the implementation of ERP systems.

The purpose of this research project is to contribute to the discussion on Enterprise Resource Planning (ERP) systems. Furthermore, the growth of businesses across the world has contributed to the increase in the use of ERP systems. The implementation of cloud based platforms for ERP systems has also made it easier for these organisations, as the infrastructure and maintenance are done by the cloud service provider. As a result of this, more and more organisations are considering cloud based ERP systems in order to improve their business processes. Microsoft Dynamics NAV is a global (ERP) solution. NAV helps organisations simplify their production, daily operations, and supply chain management. Moreover, it allows small and medium sized businesses to have more control over their finances and spending.

“Cloud computing is a model for enabling ubiquitous, convenient, on-demand network access to a shared pool of configurable computing resources (e.g., networks, servers, storage, applications, and services) that can be rapidly provisioned and released with minimal management effort” (Mell & Grance, 2011). This study explores the use of ERP solutions (including cloud based ones) for small and medium sized enterprises. In particular, this study looks at the adoption of ERP systems in Hawke’s Bay, New Zealand, and the potential impact of using cloud-based ERP systems. This project will also be beneficial for other future research projects on this topic.

As part of this project, an online survey has been conducted to evaluate the use of Microsoft Dynamics NAV as a specific example of a popular ERP system in this region. The survey participants were experienced ERP users and interested in taking the survey. The research questions were prepared according to the perspective of users in Hawke's Bay New Zealand.

## Research Questions

*Question 1: Why are cloud-based ERP systems more feasible or desirable than locally hosted ERP systems?*

There are several reasons for choosing a cloud-based ERP implementation. First of all, some industries are experiencing challenges, and are trying to implement ERP solutions in a most cost effective way. A recent survey of Chief Financial Officers around the world conducted by Deloitte shows that restrictions in Information Technology (IT) budgets have worked as a major driver for organisations to switch to cloud computing, as mentioned in an article by George (2016). In the same article, it is also reported that organisations that have moved the majority of their systems to the cloud have achieved an average 15 percent reduction in their IT spending. Moreover, cloud-based ERP systems require less maintenance and testing, with the assistance of cloud vendors for backups and support. Organisations with cloud-based applications need less guidance from consultants, as well as less technical support (George, 2016). With the cloud, implementation is quicker, and the business does not have to worry about the hardware. Cloud-based ERP implementation can be done in four to eight months, in comparison with 12 to 36 months for an on-premises installation (Catteddu & Hogben, 2010). Cloud based systems require less internal resources, and therefore are also desirable from the perspective of the human resource departments. Furthermore, small and medium sized organisations do not have to invest money in new IT hardware if they choose a cloud based ERP (Catteddu & Hogben, 2010).

Since cloud solutions do not require physical infrastructure, an ERP system helps an organisation by enabling low implementation costs. Then, organisations can focus on the other business and user issues. The cloud ERP vendor does the regular upgrades for the system. To make the businesses more productive there is a need to streamline procedures, operations, and other controls (Bourgeois, 2014). Cloud-based ERP systems can provide access to employees with mobile devices. Developers can focus on the interface, and other matters of importance for end users.  A cloud-based ERP system provides transparency in network and application utilization (Delgado, 2010). With a cloud-based ERP system, organisations can allow employees to use their own devices. In conclusion, a cloud-based ERP system can help an organisation reduce costs in different ways.

*Question 2: What are the difficulties in selecting and implementing ERP systems?*

Before coordinating business capacities, managers must consider some essential issues that will help them to choose the right ERP solution for their businesses. These include hierarchical changes, individual job responsibilities, and the different ways to deal with ERP implementation. First, managers must take into account the business's vision and corporate targets. The managers must also consider if they want to carry out a business transformation, and whether they have the capability to roll out any improvements in the business structure and operations. Management needs to select the key business functions and how these could be executed using the ERP solution. The changes to the business functions should not be extreme, otherwise it may cause a reaction from the users (Helo, Anussornnitisarn, & Phusavat, 2008). For small and medium sized organisations, it may be difficult to manage system adoption, depending on the organisational culture. The organisation must be adaptable for the new system to be successful. New hardware and additional technical experts may also be needed for software customization and implementation (Stackpole, 2014). Therefore, organisations need to consider future developments and innovations when using their ERP in an e-business capacity (Helo et al., 2008).

People-related issues, for example, corporate management approach, can play an essential part in the ERP adoption. Research shows that top executive backing is crucial to the accomplishment of any system implementation (Skilton & Director, 2010). Official meetings comprising of top executives are carried out to deal with these IT activities. It is especially important for the individuals who know a lot about the operations to be included in these meetings. Everyone should be instructed about the specific ERP product. Such instruction should include the essential ideas behind ERP systems, and also actual screen shots of the modules (Skilton & Director, 2010). During these instructional courses, it is imperative to talk about the administrative issues and to assemble a fundamental comprehension of the ideas for implementing an ERP system.

Managers must assume responsibility of the implementation process at all times (Skilton & Director, 2010). They should also direct the reengineering of the key business process, reassign work responsibilities, and rethink social connections. They should also consider how to deal with the ERP product vendors and outside experts.

ERP use requires organisations to reengineer their key business process, improving old methods for conducting business, rethinking work responsibilities, and rebuilding the organisation. For major multinational organisations, the ERP systems must be customized to address worldwide issues where diverse countries have distinctive methods and rules. Often, ERP solutions have a Western world view in them, which can be an issue for other cultures.

Organisations need to have some formal meetings and surveys in order to make decisions. The vendors also carry out some evaluations to make sure that this ERP is the right choice for the business. They may also explain their clients about the activities to be carried out when upgrading the ERP systems. For example, the following are some steps to be kept in mind when implementing ERP systems for businesses (Processpro, 2016):

- Analyse Project Scope
- Develop a Strategy
- Define the Standards
- Test, Validate, and Reconcile
- Staff for the Task

ERP vendors have defined their ERP systems around standard business processes, based upon best practice (Parpia, Hashemi, Bennett & Sekuler, 2015) Different vendors have distinctive procedures. Firms are compelled to adjust their procedures rather than adjusting the ERP system to the current procedures. It is pivotal that organisations conduct an investigation before selecting an ERP vendor. The research should cover all operational procedures.

Businesses need to choose the best ERP vendor and the most suitable software product. ERP implementation is more difficult and sensitive for small to middle sized organisations, business rules, information semantics, approval procedures, and choices. They need to start by choosing which modules to introduce, then modify the system to accomplish the most ideal fit with the organisations' procedures.

Most ERP systems have the adaptability of implementing selected features and not others. Some basic modules, for example, customer relations and accounting are embraced by most organisations undertaking ERP systems. Other modules, for example, human resources, are not required by some organisations. An administrative organisation for instance, will not likely need a module for product assembly. As a rule, the greater number of modules chosen, the greater the disadvantages, costs, risks, and changes. Consulting costs are important because many ERP products are not ready to be used off-the-shelf. At present, many companies are offering consulting services for ERP implementation. Usually, this includes the whole ERP implementation process: preparation, testing and delivery. The cost of ERP systems depends on the number of ERP client licenses.

*Question 3: Evaluate a commercial ERP solution, i.e. Microsoft Dynamics NAV (rather than an open source solution e.g. webERP) based on its usefulness for NZ businesses.*

The significant point of preference of open source ERP solutions is the greater customizability (Erturk, 2009; Wang, He, & Wang 2012). On the other hand, there are drawbacks of open source ERP packages, e.g., possibly fewer customers and less technical support. While providing normal ERP functionalities, every open source ERP is unique. Among open source ERP systems, webERP appears to one of the best for four basic reasons: compatible with distributed computing, standard ERP functionalities, ease of setup, and customer communities. webERP also has less demanding system requirements. On the other hand, webERP has not become popular, especially in New Zealand (including Hawke's Bay). The reason for this maybe a lack of IT professionals who are aware or experienced with that open source solution (Fougatsaro, 2009). Commercial ERP solutions tend to have a strong track record, and market share. It may be a safer choice for some organisations to adopt a popular ERP system like Microsoft Dynamics NAV and SAP, even if they cost more. After an initial informal investigation, it has appeared that Microsoft Dynamics NAV is relatively popular in Hawke's Bay, among agricultural, viticulture, and food-related companies. Therefore, this project proceeded to create a survey to explore the use of Microsoft NAV by these types of regional companies.

## Introduction

The Enterprise Resource Planning (ERP) system is "an enterprise information system designed to integrate and optimise the business processes and transactions in a corporation" (Moon, 2007). The ERP system is industry-driven and is a solution for every organisation with specific goals. This research has various objectives. First, this study will be significant to other researchers interested in general ERP systems as well as cloud-based ERP. Secondly, this study will present a recent review of papers and reports in journals and periodicals from 2006 to 2016. This review contains approximately 40 sources. Section 4 contains the literature review. Section 5 explains the methods. Section 6 portrays the findings. Section 7 then clarifies the limitations. The last two sections are the conclusions and recommendations.

"An ERP system enables an organisation to integrate all the primary business processes in order to enhance efficiency and maintain a competitive position" (Addo-Tenkorang & Helo, 2011). However, without successful implementation of the system, the projected benefits of improved productivity and competitive advantage would not be forthcoming. The ERP system stands for enterprise resource planning. Enterprise Resource Planning (ERP) is an enterprise-wide information system that integrates and holds all the business processes in the entire organisation (Addo-Tenkorang & Helo, 2011). ERP systems have become a vital tool for all the businesses in today's competitive business environment. The ERP system is an enterprise information system designed to integrate and enhance the business processes in an organisation.

The ERP system eases the smooth flow of communal information and practices across the entire organisation. Furthermore, it improves the performance of the supply chain and reduces the cycle times. However, "without top management support, having appropriate business plan and vision, re-engineering business process, effective project management, user involvement and education, organisations cannot hold the full benefits of such complex system and the risk of failure might be at a high level" (Addo-Tenkorang & Helo, 2011).

Smadi (2007) indicated that implementing ERP systems can raise benefits for businesses, for example reducing cycle time, improving flow efficiency, and quickly generating monetary information. It was in the start of the 1990s when ERP systems were first introduced (McGaughey & Gunasekaran, 2007). According to the business reaction, it was a great product. Yet, from the perception of system developers, it was a challenge to implement (Almgren & Bach, 2014). The ERP system is much more than integrating diverse subsystems into one gigantic system. It is a system through which businesses consolidate their information assets (Almgren & Bach, 2014). However, everything innovative carries new challenges. ERP may not only create difficulties for developers, but for business end users as well (McGaughey & Gunasekaran, 2007).

The overall development of ERP systems has allowed all organisations to maintain their business operations and transactions. ERP is basically an information system that combines diverse subsystems into one system. Information systems help in increasing the efficiency of all business operations (Parto, Sofian, Saat & Mohamed, 2016). This process is called integration, whereby subsystems are integrated into one system. For example, an organisation has three leading information systems. The first system handles human resources; the second system handles finance; and the third system handles manufacturing (McGaughey & Gunasekaran, 2007). ERP integrates these three subsystems into one system that shares data among these subsystems organisations (McGaughey & Gunasekaran, 2007).

ERP systems allow all organisations possible growth. In order to reach their goals, organisations are increasingly implementing the ERP systems. A significant growth in use by many types of businesses has been observed in recent years, (Dey, Clegg &

Bennett, 2010). The worldwide ERP business sector's incomes were estimated in 2008 as $61 billion, and in 2010 as $65 billion (Parto et al., 2016).

In terms of the expertise required for ERP system implementation projects, the majority of the decision making is done in the initial implementation phase of the ERP, for example, choosing an appropriate centralised or distributed ERP system. The use of the ERP systems may influence the effectiveness of information systems through improved management decision-making processes, internal controls, and improving the quality of financial reporting, and facilitation of transaction processing (Sajady, Dastgir, & Hashem, 2008). Additionally, an effective information system is anticipated to increase the quality of choices, and improving the organisation's performance.

Successful organisations need the right ERP solution to help their main business processes and make smarter and faster decisions, and ensure they make the most of their assets and resources.

## Literature Review

In order to understand the adoption of ERP systems, the history should be reviewed. ERP solutions date back to the 1960s when the early accounting and inventory systems were introduced (Elragal & Haddara, 2012). Monitoring operational expenses was the main competitive thrust in the 1960s (Jacobs & Weston, 2007). Consequently, manufacturing plans became more product-centered based on high-level volume production, the minimization of costs, and presuming solid financial requisites (Jacobs & Weston, 2007). According to Motiwalla & Thompson (2008), enterprise resource planning systems are early generation enterprise systems that target the integration of data and to provide support to the organisations main functions. The development of ERP software from 1960's to today has been affected by other major IT inventions (Plex, 2015). The development seen from minicomputers to the cloud is helping the organisations to assemble their businesses.

Most small- and medium-sized enterprises use basic business software to manage their daily operations (Almajali, Masa'deh, & Tarhini, 2016). Eventually, they consider changing to an ERP system. However, implementing ERP system successfully is costly and complex, and frequently shows high disappointment rates if the ERP system does not readily align with the company's business requirements or their social environment (Almajali, Masa'deh, & Tarhini, 2016).

ERP products are designed to help organisations work more efficiently (Plex, 2015). The first ERP systems were similar to Manufacturing Resource Planning (MRP) systems and were used by other types of organisations. MRP (material requirements planning) and ERP (enterprise resource planning) were implemented to manage the operations of these organisations. The two main concerns in past years were business procedures and accounting. ERP solution offers some products like Oracle, JD Edwards, SAP and more. As the organisations grow, finance teams must give guidance to the business and be able to analyse data, rather than storing data and assembling it. Microsoft Dynamics Great Plains, helps organisations to maintain and perform different functions (Awsi, 2013). Human resource management, financial management, operational management and assembling are the elements of an ERP system (Awsi, 2013). With the help of an ERP solution, the business processes can be handled in different ways. Great Plains has a number of principles (Microsoft, 2016a):

1. Ability to deal with diverse portions of business' processes
2. Noteworthy, understanding how well your organisation is running, and find out the upgrades needed
3. Adaptable group choices to provide more efficiency
4. Coding should be simple and useable
5. Enhanced business selections may prompt long-drawn-out income and boundaries

ERP system needs to provide more adaptability to users and minimal effort in the future. Cloud computing can help organisations to implement and maintain ERP systems more easily. "Cloud computing is a model for enabling ubiquitous, convenient, on-demand network access to a shared pool of configurable computing resources (e.g., networks, servers, storage, applications, and services)" (Mell & Grance, 2011). Cloud computing is seen as an evolutionary progression in computing during the last decade (Marston, Li, Bandyopadhyay, Zhang, & Ghalsasi, 2011). As a result of cloud computing, managers can achieve greater productivity at less cost, bringing about genuine enhancements in assembling operations. ERP users have many different demands; some of those changes can be outsourced and done by cloud based service providers.

The business procedures and their data requirements as well as the use of cloud-based ERP systems have changed operational procedures. Moreover, implementing ERP systems is complex because expenses may be too high or the system may be met with potential disappointment (Wailgum, 2009). ERP systems help in bringing the changes in a business, according to every business needs. Additionally, ERP systems implementation is done on the basis of business procedures for SMEs and expansion plans. The standard method is to outsource the installation and configuration to other businesses, especially consultancy companies (Pollock & Williams, 2008).

In an organisation using ERP systems is a way to increase the quality and productivity of various business processes. ERP system helps many commercial businesses and several practice areas in a synchronized manner, endeavouring to automate operations of the accounting, management, maintain records, user relationship concerns, stock control monetary and, HR and other useful areas in an organisation. Research by Abdullah, Albeladi and Atiah (2013), for example, looked at the acceptance rate and the impact of ERP systems by various types of users specifically; functional users, external users, and technical staff. Relatively little prior research on ERP systems is another principle driver for this research project.

ERP systems implementation has focused on some of the major areas, for example, operational management, key performance indicators, and quality assurance. Most existing studies report just on the 'positive effects' (advantages) of ERP systems in one hierarchical level and frequently they have been directed just inside small number of organisations. Moreover, Information Technology (IT) related papers in recent years have looked at the advantages and disadvantages of IT, from the perspective of how information is captured and used by those systems (Hosseini, Sommerville, & Sriram, 2011). It is important to protect the privacy of information at different levels, i.e. the economy, the particular industry, and the individual firm level. The significance of information governance at various levels within the organisation has also been a critical issue in recent years (Hosseini et al., 2011).

A research study by Sedera, Gable and Palmer (2002) for example, looked at the effects of ERP systems, particularly SAP, among ERP users in Queensland, Australia. According to the study, for some businesses, the method they used to choose their ERP system could have been a disadvantage in itself.

ERP systems help organisational departments to work together:

- Integration among various divisions to guarantee appropriate communication, effectiveness and usefulness (Feurer, 2007).
- Permits control of cross-functional business processes and minimises the so-called islands of information.
- Reduces the risk of loss of data by joining various authorisation and security models into a singular structure (Farquhar& Hill, 2013).

Some security components are integrated into an ERP system to protect the system against security threats and attacks. In addition, there may be insider threats from dissatisfied internal users. An ERP system can improve accuracy, consistency and security of data. Restrictions to data can also be improved. ERP vendors are additionally moving with different sorts of data security tools. Compatibility Issues with ERP modules lead to issues in integration of modules. Companies associate different vendors to implement different ERP modules, based on their competency. It is very necessary that there is a way to handle compatibility issues.

“Even though the great recognition and acceptance of ERP Systems in organisations, some criticisms have been directed to these types of systems, whether from a technical standpoint or from a business perspective” (Azevedo, Romao & Rebelo, 2012). Implementation of ERP systems is complex and very costly. ERP systems are sometimes seen as slow and inflexible. This may be one of the reasons for user dissatisfaction. Networks need to perform at a high capacity for applications to work effectively.

An organisation can achieve their targets, but inaccurate data will reduce the reliability of the ERP applications. When an ERP system is implemented, transaction incidentals are also high for all the ERP users (decreasing flexibility and dynamic change in the organisation). The organisation boundaries can raise difficulties in accounting, business services, and employee support. Some organisations may have many separations with isolated, autonomous resources, tasks, management hierarchies, and more. The ERP systems are also critical for decentralized organizations with diverse business processes and systems. ERP systems integrate the information in one spot. This can expand the risk of losing significant data in the instance of a security break. Once a system is developed, costs are high for any of the organisations, reducing flexibility and control at the corporate level (Sundar, 2013). The system may be too confusing for the basic needs of some customers. ERP Systems put the data in one place. This can increase the risk of loss of sensitive data if there is a security breach (Ngai et al., 2008).

It is a logical decision to implement ERP systems for SMEs in the cloud since it might offer some noteworthy accessibility and flexibility (Mahmood & Hill, 2011). Whether implemented locally or on the cloud, some of the ERP solutions available are:

**Brightpearl:** Intended for medium sized multichannel retailers, this cloud ERP system empowers them to deal with their core business by joining orders, stock, client information, bookkeeping and reporting into a single retail management system (Brightpearl, 2016). Brightpearl's Commerce Acceleration Platform conveys stock and revenue data, SKU and channel productivity, and user behaviour.

**OfficeBooks:** OfficeBooks is a simple to use, business management application. OfficeBooks can help run the business effectively. The application coordinates all the key procedures of any business (Microsoft, 2016b). Best of all, OfficeBooks is hosted in the cloud so it is accessible anyplace; at the workplace or at home.

**webERP** is an online office and business management system that needs just a web-database to access the information. webERP is a pioneer of distributed computing where the application can happen anyplace on the Internet and be deployed to the business through the cloud. When joined with an external intelligent desktop Point of Sale system, it can also become the central component of a multi-branch retail administration system. A completely coordinated webSHOP is additionally accessible as an external add-on (webERP, 2013). webERP is as an open source software, and is available for free download (with PHP source code), and can be customised with the user's own particular elements and configurations as required (webERP, 2013).

**Microsoft Dynamics NAV:** Coordinated business management system that computerizes money related, client relationship and inventory network forms. Microsoft Dynamics is a line of coordinated, versatile business management systems that empower the business to settle on critical business choices with more noteworthy certainty. Microsoft Dynamics works like and with well-known Microsoft programming facilitating appropriation and decreasing the risks typical of using a brand new system. These systems robotize and streamline budgetary, client relationship, and inventory network forms in a way that can help drive business achievement. Microsoft Dynamics NAV is a universal ERP system that provides small to mid-sized organisations a prominent control over their financials and can help with their whole supply chain, production network, assembling, and operations (Microsoft, 2016c). Microsoft Dynamics Microsoft Dynamics Nav and Office 365 deliver the tools to manage the business and help organisations to achieve more. According to Microsoft (2016c), the following modules are available:

- Financial administration and bookkeeping, including managing money
- Supply chain, assembling, and operations, including tracking and dealing with production, stock, requests, and suppliers
- Marketing, deals, and administration, including contacts and contracts
- Project administration, including milestones and deliverables
- Business intelligence and reporting
- Multi-currency processing abilities

As seen in the literature review, there are many ERP vendors providing similar benefits for organisations. Microsoft NAV is a complete, recognized, and trusted solution for New Zealand ERP users and works in synergy with Office 365, which is popular in New Zealand (Microsoft, 2016c). This project focuses more on this particular product (Microsoft Dynamics NAV) and the local users' experience with it and their future plans for their systems.

## Methods

This study has followed online surveys because it is the least costly method for collecting information from individuals. The current Microsoft Dynamics NAV users are also experiencing cloud based systems. Therefore, it may be interesting and more familiar for them to do an online survey using typeform.com. This approach will also allow more time for the users for responding. Online surveys have become common because they are fast. With online electronic surveys, the questions can be designed in different ways, for example, multiple choice. This way, it is not difficult to communicate with and collect information from the organisations and individuals who are using Microsoft Dynamics NAV.

The exploration will examine the client's experience (related to the third research question) and will discover some of the difficulties in implementing cloud-based ERP systems (related to the second research question). The investigation will characterize the nature of cloud based ERP systems (related to the first research question) and some ideas for picking the right ERP system for businesses. Furthermore, Section 7 will address the first research question. Section 8 will also address the first and the second research questions. Section 9 will have more discussion on the third research question.

After getting a deeper understanding about ERP systems from the literature, this exploration moved to creating an online survey for investigating the effects of Microsoft Dynamics NAV. Another objective of this research report is to understand the implications of ERP systems for small and medium sized enterprises (i.e. fewer than 100 employees). Some of the questions allowed open-ended (qualitative) responses, which have also been read and analysed in this report.

Research Design: the writing of the survey questions has been influenced by the literature review, including journal articles about cloud-based and ERP systems. A review of the Microsoft Dynamics NAV software has also helped to interact better with Hawke's Bay NAV users. Other parts of this research include conference presentations, professional meetings, reports, interviews, as well as peer-reviewed journals and articles. Prior to the survey, an interview (or discussion) was conducted with an experienced local Microsoft Dynamics NAV user, Ian Purdon (personal communication, May 25, 2016). This has provided a much better understanding of the local NAV user base and their needs, prior to the survey.

The Microsoft Dynamics NAV users in question include companies that use the software and their main NAV users and managers. The method of recruitment is by approaching members of the already existing Hawke's Bay Local NAV Professional Users' Group. The participants have been sent an email invitation.

**Ethical Considerations**

The research survey questionnaire is anonymous. The respondents were volunteer expert users in the Hawke's Bay Region. Consent was obtained from the users before their participation in this research project. As a result, official ethics approval has been obtained from the Eastern Institute of Technology Research Ethics and Approvals Committee. The summarized statistical results of this survey will also be shared with the users.

## Findings

Data collection from Dynamics NAV users was carried out through an online survey. The survey was conducted on typeform.com, and investigated the use of Microsoft Dynamics NAV as an ERP solution in the Hawke's Bay Region. Some of the major findings will be mentioned below in this section. The survey consists of a total of 25 subjective and objective questions. The total number of respondents were five. The users who responded seem to have been well engaged with the survey. This is an acceptable number for an exploratory study which involved not just any users, but only those that were well experienced with Dynamics NAV ERP systems. The results provide an overview about the usage of ERP systems in the Hawke's Bay Region. Figures 3 through 8 below show the results as bar charts.

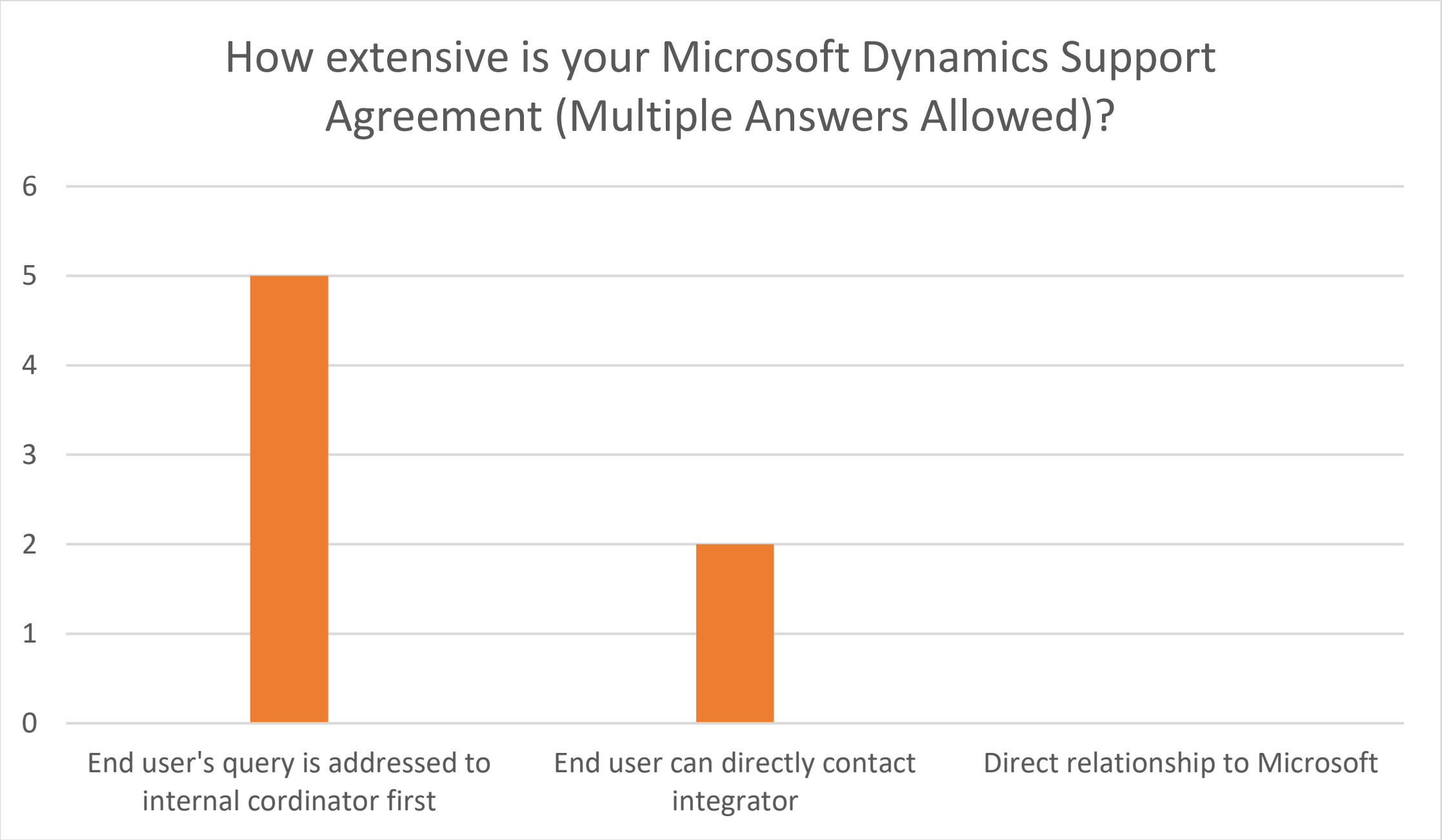

*Figure 1: User Support*

As seen in Figure 3, most of the queries are handled internally. In the case of integrators, this survey has identified three major NAV integration and support companies active in New Zealand: Acumen Consulting Limited, Fujitsu NZ Limited, and UXC Eclipse.

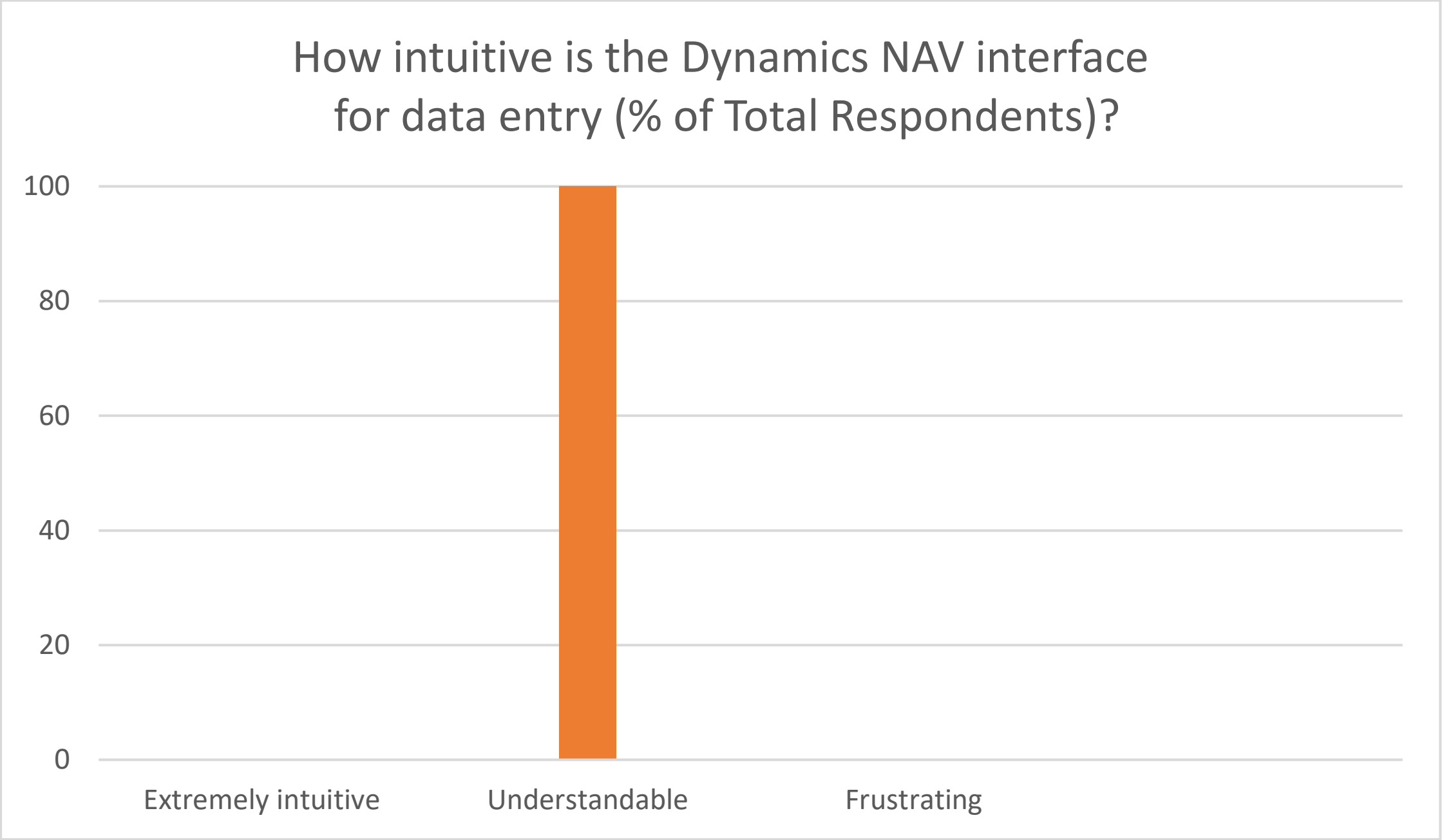

*Figure 2: User Interface*

The above figure 4 illustrates how intuitive is the Dynamics NAV interface for data entry in the Hawke's Bay region. All the users agreed that it is understandable.

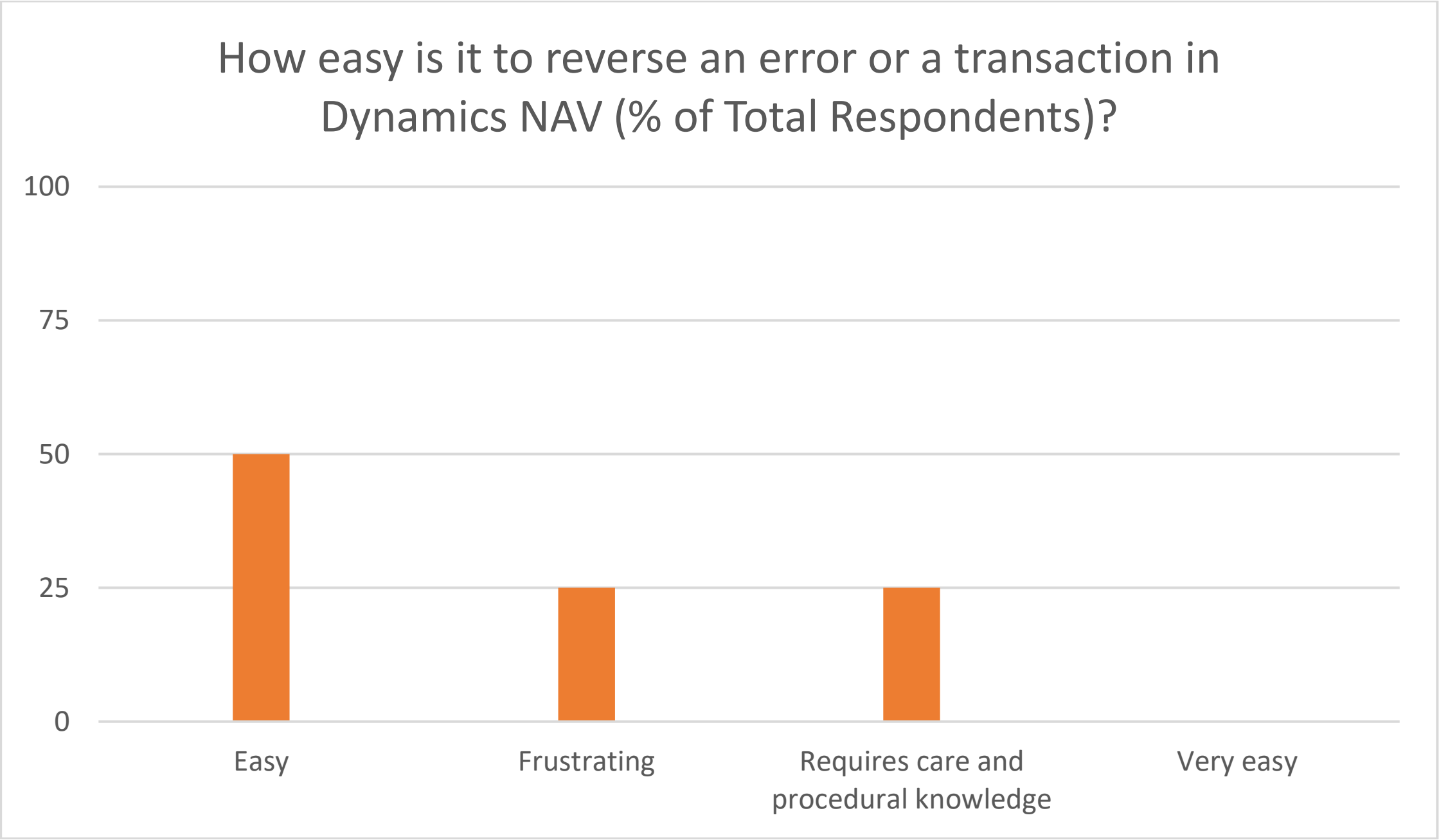

*Figure 3: Ease of Use*

In figure 5, out of five users, two users found it easy to reverse an error or a transaction in Dynamics NAV. One user said that it requires care and procedural knowledge and the other user found it frustrating.

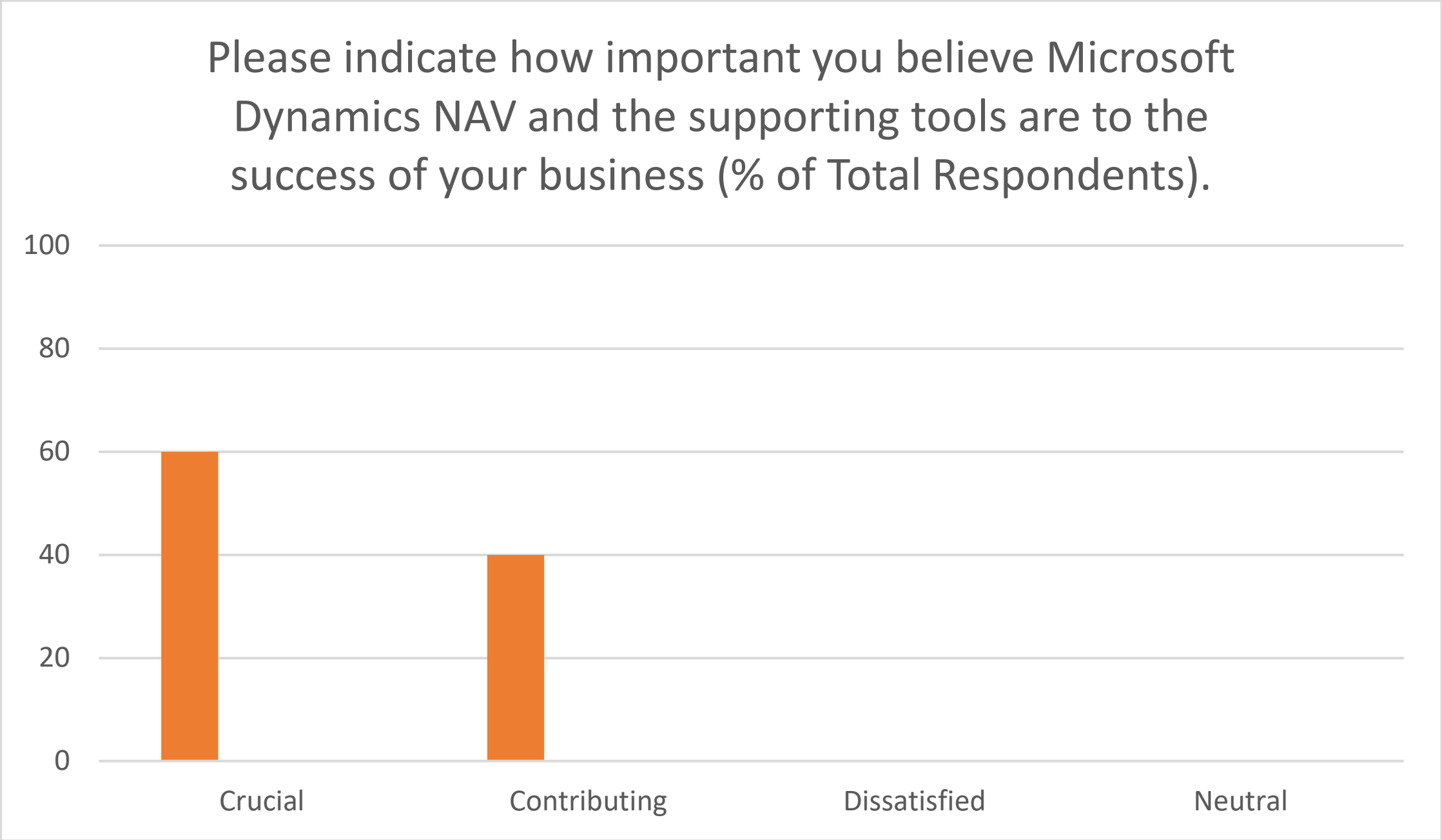

*Figure 4: Supporting Tools*

The above bar chart depicts that most users believe Microsoft NAV and the supporting tools are crucial to the success of business whereas only one user finds it contributing.

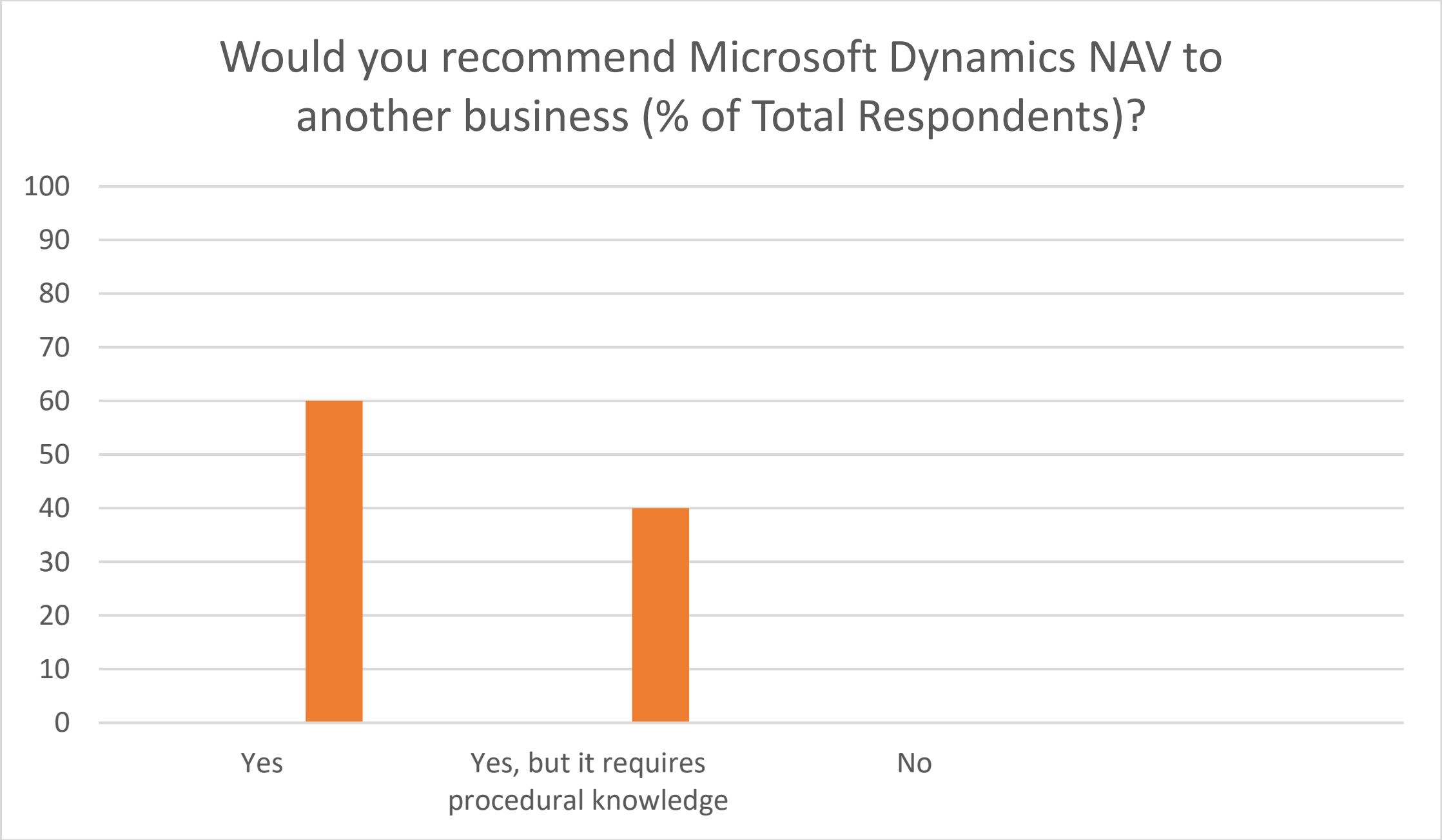

*Figure 5: Recommendability of Microsoft NAV*

Figures 7 and 8 describe that 80 percent of the users would recommend Dynamics NAV to another business. While only 20 percent users of Dynamics NAV may recommend it to the others although it requires procedural knowledge.

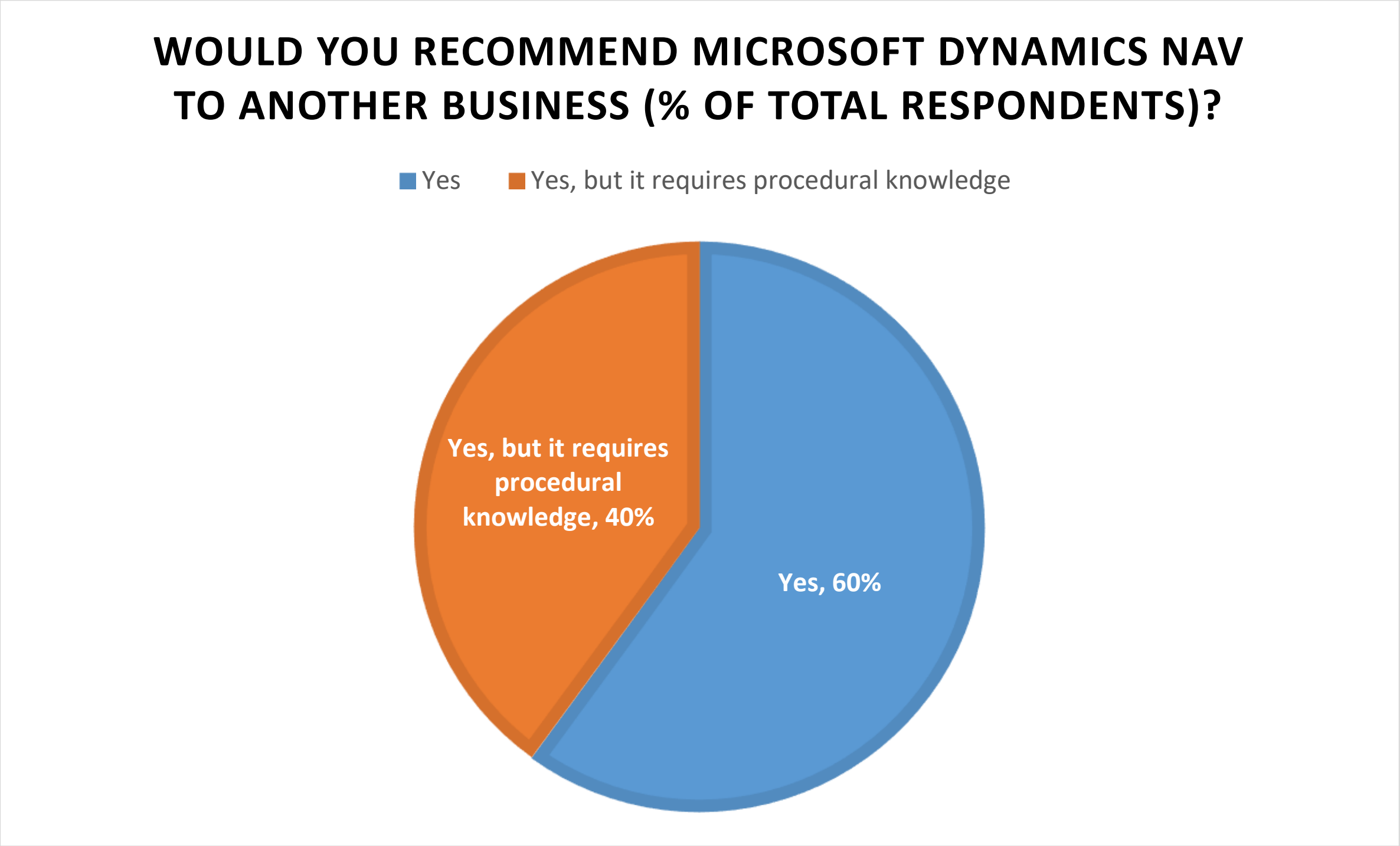

*Figure 6: Recommendability of NAV (Pie chart)*

The ERP users had a variety of answers for the questions, which makes this survey interesting to analyse and repeat in the future. For example, the participants ranged from those planning to implement it later this year to other companies that have been using ERP systems for seven or ten years. This includes versions of NAV 2009 to NAV 2016. Some of these local companies believe that the EPR product was relevant to their business, whereas some of the participants were influenced or directed by their parent and other companies.

In almost all of the companies that responded to the survey, the new user training is provided by the managers in the same company who have learnt this product before. Similarly, out of five companies four companies have indicated that they will continue to use Microsoft Dynamics NAV in the future. Therefore, the results of this survey are very helpful for these companies because they plan to use the NAV ERP in the future.

The survey results show that some of the companies have already upgraded a few months ago and the others are planning to upgrade their NAV system within the next 18 months. The users would always like new features, and greater compatibility and integration with other software packages. When users would like a new feature, their queries are addressed to management first before any technical action.

There were a number of questions that asked users to describe the modules of Microsoft Dynamics NAV that they are using. The findings here are interesting as each company has different preferences for their Microsoft Dynamic NAV ERP system. Throughout the Hawke's Bay region, the modules currently in use include: logistics, advanced warehousing, manufacturing, financial management, sales and marketing, purchase, warehouse, document capture (in combination with Continia), expense management, general ledger, accounts payable, accounts receivable, FA, sales, inventory, meat product management, quality management, winemaking, vineyard management, and electronic data interchange. Hawke's Bay Microsoft Dynamic NAV users are also working with additional third-party report writing tools for business intelligence, for example, Power BI, JET Essentials/Enterprise and COGNOS/TM1. From the findings, we can see that even in a regional setting, an ERP system can be very complicated with multiple modules and third party tools working together.

The challenges and issues of current Microsoft NAV ERP systems can be seen from the Hawke's Bay NAV users' comments. One user mentioned "one badly designed/customised data entry aspect." The same user also reported experiencing "restriction on true multi-user journaling," and "sub-optimal initial configuration limiting its use for planning/scheduling" as their particular end user issues. Another user said that "getting users to test it properly when changes are made" is necessary. Therefore, testing is an essential phase during the implementation.

The responses to the remaining questions were qualitative. Therefore, these long responses are presented on the following pages in detail rather than summary charts.

| **How long have you been using NAV?** | **Why did your company choose Dynamics NAV?** | **Which company provides the majority of your NAV support / Development?** | **How do you arrange new user training for NAV inductions?** | **Which version of NAV are you using (for how long)?** |
|---|---|---|---|---|
| 2 years | Hobson's choice - basically weren't given the option by our parent organisation | UXC Eclipse | N/A yet as we have not commenced user training. Sessions to date have been small internal groups, hosted by Acumen Consulting on-site. | 2013r2 - 2years |
| Not yet implemented - due 30 September | Microsoft product has familiar look and feel, good support and maintenance offering, future-proofed, solution met our requirements | Acumen Consulting Limited | 2 hours provided gratis by NAV Integrator for generic NAV interface etcetera, then responsibility of department manager for on the job familiarity | We are implementing NAV 2016 on premise |
| 10 Years | It was a reasonable fit for a mid-sized company and it contained a vertical which assisted in processing meat industry transactions | Fujitsu NZ Limited | Generally new users are trained by existing users of the system | NAV 2016 (using for 6 months) |
| 7 | Winemaking | UXC Eclipse | Core team members do it - already trained in tasks | 2009 |
| 3.5 Years | We use Vinpoint which is a winery specific derivative of NAV | UXC Eclipse | In House | 2009 R2 |

| Do you envisage staying on the NAV platform in the foreseeable future? | Are you considering an upgrade to a new version? Why or why not? | Please describe which modules of Microsoft Dynamics NAV you are using. | How intuitive is the Dynamics NAV interface for data entry? | Do you use multiple currencies in Dynamics NAV? |
|---|---|---|---|---|
| My understanding is that this is being re-evaluated as part of Group-wide SAP roll-out within 2-3 years | no - cost and see prior answer i.e. unknown future | all modules including logistics, advanced warehousing, and manufacturing | Understandable | Yes |
| Yes | No - we are going-live with 2016 (latest version) | Financial Management, Sales and Marketing, Purchase, Warehouse, Document Capture (Continia), Expense Management | Understandable | Yes |
| Yes | Just upgraded a few months' back | GL, AP, AR, FA, Sales, Inventory, MeatPro | Understandable | Yes |
| Yes | Yes | Financial Management, Sales & Marketing, Purchase, Quality Management, Winemaking, Vineyard Management, EDI integration | Understandable | Yes |
| Yes, the cost and effort of change is prohibitive | Yes, in the next 18 months to take advantage of new features and compatibility, particularly around integration with other software packages | Financial Only | Understandable | Yes |

| **Do you use an additional / third-party report writing tool (BI) with Dynamics NAV?** | **If you answered yes to Question 12, please name which reporting tools you are using:** | **How easy is it to reverse an error or a transaction in Dynamics NAV?** | **Would you recommend Microsoft Dynamics NAV to another business?** | **Do you have any suggestions for improving NAV products/services?** |
|---|---|---|---|---|
| Yes – Which Tool(s) | | Frustrating | Yes | never liked the advanced warehousing add-on feel i.e. should be more intuitive; the critical fields in a form should be easier to highlight for users i.e. not require programmatic definition |
| Yes – Which Tool(s) | We WILL but still in analysis phase - Power BI, JET Essentials/Enterprise | | Yes | Too early in implementation to be able to provide an answer |
| Yes – Which Tool(s) | JET Essential and JET Enterprise | Easy | Yes | |
| Yes – Which Tool(s) | Jet Reports | Requires care and procedural knowledge | Yes | |
| Yes – Which Tool(s) | COGNOS/TM1 | Easy | Requires care and procedural knowledge | |

| Please indicate how important you believe Microsoft Dynamics NAV and the supporting tools are to the success of your business. | What are the challenges and issues of the current Microsoft NAV Implementation? | How many concurrent users does the current Microsoft Dynamics NAV licence support? | How customized are the NAV user profiles? Do you use multiple NAV profiles? | What database technologies support the NAV environment? (SQL Server, and version…) |
|---|---|---|---|---|
| Contributing | one badly designed/customised data entry aspect; restriction on true multi-user journaling; sub-optimal initial configuration limiting its use for planning/scheduling | 32 | some good profiles, but not widely implemented or tuned - lack of designer skill/time | SQL Server 2012 |
| Crucial | Solution design to integrate core operational system into NAV | 41 concurrent users | Cannot comment yet | 14 |
| Crucial | Getting users to test it properly when changes are made | 25 | Not customised at all. Only use one profile. | 12 |
| Crucial | | 30 | Very - yes | 14 |
| Contributing | | 5 | No - all as super - the controls around limited user access are too complicated | |

| Did you consider hosting and running NAV on the Cloud? Will you reconsider this in the near future? | What would be your main concerns with cloud hosted NAV? | Do you allow remote access to NAV? | Additional Comments: |
|---|---|---|---|
| no; and not for the foreseeable future | bandwidth; uptime; data extraction for local reporting | Yes | NAV has significant upsides, just awaiting in-house maturity and resources, and potentially take-up of upgrades to enable new features |
| Yes, we did but was prohibitively expensive for licencing. We will reconsider in the future. | Performance | Yes, we will do | As we are in the early stages of the implementation I have not been able to answer all questions. I hope this feedback is sufficient with what I have provided. |
| No not at this point | Speed and security of data | Yes | |
| We will consider in the next 5 years possibly | Speed and cost | Yes | |
| No, rural based facility so internet is restrictive | as 24 | Yes | |

The conclusions drawn from these results can be found in Section 9. Sections 7 and 8 are based on literature review. Section 7 addresses the first research question. Section 8 addresses both the first and the second research questions in this project. Section 9 will address the third research question further.

## Difficulties during and post ERP Implementation

Enterprise Resource Planning systems are related to Material Requirements Planning (MRP) systems which were exclusively centred around production and materials (Panorama, 2010). ERP systems were intended to oversee the elements of these undertakings by coordinating all business management functions, including ordering, stocking and human resource management, designing, accounting, and so forth. On the other hand, ERP did not consider any of the external stakeholders that an organisation has (Panorama, 2010). Inside an organisation, the application could support all capacities, but it may handle different communications that the organisation had with its suppliers and customers.

ERP systems are extremely complex and costly to implement. The time required for an organisation to move to an ERP system is very long. Comprehensive preparation must be done, within the normal working hours; this can cause regular business to slow down. Aside from the establishment costs, organisations charge yearly maintenance and renewing costs. These costs set up together may not legitimize the use of ERP systems particularly for organisations (small and medium sized enterprises). Since, every business has different necessities there must be a level of customization to guarantee that all the components relevant to the business are considered. This may call for changing the ERP software design structure to coordinate the business work process which is not permitted by the product sellers.

ERP systems are controlled according to industry norms and when an organisation needs to use the system, it might need to change its method of working and to coordinate these principles. This can either be valuable or can prompt the business about risks. Some organisations moderately have basic operations and the use of ERP systems may alter the current setup, hence prompting an over-designing when compared with client's requirements. There may be cases where divisions are disinclined to share information because of decreasing the adequacy of ERP systems. Many organisations keep running legacy systems and ERP systems at the same time, which results in data duplication across the departments (Manojlov, 2012). Conversion from the legacy systems in itself results in high costs.

The simple accessibility of data takes up the issue of surety. Admittance to the different functionalities must be intentionally managed to avoid unauthorized access and information loss. The time period to realize the full benefits of an ERP product is long. With time, there will be changes in the business requirements of the organisation. Since the ERP product may be updated during the implementation phase, the organisation will have to be versatile. To overcome this, organisations would need to ask the vendor for more assistance because of the future updates. The specialized support provided by the product vendors is crucial for achieving positive results.

## Implementation Phases and Risks

In the structured implementation approach, the early phase in choosing an ERP system is to obtain a planned path to adopt the business procedures (Rogers, Sharp, Preece & Tepper, 2007). Everyone desires to determine the plan for common requirements; welcome to the report; how potential vendors are selected; the purpose of presentations and the process for selecting the ERP vendor (Rogers et al., 2007). Businesses make choices between a cloud-based ERP and an on-premises solution.

Next, presentations by experienced vendors are important for the business to change. Anyhow, it is also vital that there is a great amount of willingness on the part of the clients (Rogers et al., 2007). This way, vendors can engage their clients better during the presentations.

Thirdly, choosing an ERP system is a hard decision to make, which has big financial implications (Rogers et al., 2007). There are two important areas to look at when the executive managers are deciding on the criteria that will be used when assessing potential vendors (Rogers, et al., 2007). Firstly, the criteria and the grading system must be concurred ahead of time before survey any potential systems (Rogers et al., 2007).

The selection on the system must be prepared by all stakeholders in the organisation (Rogers et al., 2007). It requires top management initiative and investments that includes every office within the company (Rogers et al., 2007). Representative of all the users should:

- Be involved in undertaking the initial phase where the basic development process is determined
- Go to the vendor demonstrations
- Have their say in the short-listing and choice of vendors

Protection and privacy risks are mapped to be among the top concerns over cloud-based ERP (Marston et al., 2010). According to the analysis, those security threats of cloud ERP choice are ordinarily more basic for significant endeavour's than for SMEs.

Aside from security issues, implementation risks of cloud-based ERP are related to compliance with data privacy regulations (Kim, Kim, Lee, & Lee, 2009).

Various cloud-based ERP structures have noticeable constraints on interoperability with locally hosted applications and linking into existing application portfolios and IT establishments (Karabek, Kleinert, & Pohl, 2011). The cloud-based ERP may not allow extensive customization and complex integration with external organisations and systems. Meanwhile, one of the most often misinterpreted issues with most ERP systems is the acceptance that they normally require extensive customization for SMEs and medium sized enterprises.

While customising an ERP system, the organisation continues through the extended risk of high requirement by an ERP system user. The cloud-based applications consistently stand up to additional difficulties in consenting to the data, and environmental models as these bearings are generally made without the appreciation of the characteristics of distributed processing (Kim et al., 2009).

As an eventual consequence of outsourcing the critical number of IT reinforce, organisations lose some imperative IT capacities and IT division's resistance towards various levelled changes (Zhang, Wuwong, Li, & Zhang, 2010). Cloud based ERP systems may focus on network issues rather than business requirements. Their major goal is to live up to the back-end IT computing and computer memory needs of the organisations (Scavo et al., 2012).

It may sometimes be hard to depend on Service Level Agreements (SLAs) for issue resolution between the cloud customer and the cloud provider (Kuyoro et al., 2011). For example, these SLAs in general do not cover such perspectives as security and privacy (Rong, Nguyen, & Jaatun, 2013).

The risk management plan in an IT project is everyone's responsibility and it should include the sponsors, customers, all project team members and all other stakeholders. The risks involved in an ERP project are classified in six different categories; executive risk, project risk, functional risk, resource risk, organizational risk and technical risk.

ERP systems manage internal and external data that upgrade the stream of data and focuses on further business processes. The risks can worry the organisational system and effect the business procedures after the go-live date when the ERP system is completely operational and accessible to end clients (representatives, lower level managers, and more). The extent of this research report investigates both pre and post go-live issues.

## Conclusions

It is obvious that New Zealand has been influenced critically by developments in IT. The advantages and challenges of ERP systems require a thorough evaluation. Other research studies focused on specific issues whereas this study has tried to give a general clarification about the impacts of ERP systems on users in a particular geographical setting. It is important to understand the implications of the answers to each of the questions. These will provide some ideas for a sound implementation in commercial and non-profit institutions.

### Conclusions Drawn from the Survey

There were a number of questions related to the user interface. One of the questions asked "How easy is it to reverse an error or a transaction in Dynamics NAV?" Although some of the users found it to be an easy task, others got frustrated with it. In other words, one of the users said that it requires carefulness and advance knowledge of the application.

Some of the participants provided feedback for improving NAV product/services. One user has specifically stated, "never liked the advanced warehousing add-on feel that should be more intuitive; the critical fields in a form should be easier to highlight for users that not require programmatic definition”. In conclusion, user friendliness and usability are important factors in an ERP system implementation.

The users were also asked how important they believe Microsoft Dynamics NAV and the supporting tools are to the success of the business. We can conclude from the responses to this particular question that most ERP users find the supporting tools to be either crucial or contributing.

**Technical and Cloud Computing Related Findings**

The survey shows that most of these companies have more than 20 concurrent users of Microsoft Dynamics NAV. Some of the NAV users believe that the user profiles are well customized, whereas a few of the companies have not yet implemented or not yet fine-tuned their generic user profiles due to a lack of technical skill or time. This includes users with only single profiles as well as those with multiple profiles. However, all of these respondents in Hawke’s Bay allow remote access to their Microsoft Dynamics NAV system. These points highlight how important networking is for contemporary ERP systems.

The main database technology used in Hawke’s Bay in conjunction with Microsoft NAV is SQL Server (version 2012 and 2014). The CRM module is rarely used in Hawke’s Bay. Some of the companies are not yet ready to host and run NAV in the cloud; whereas some of them are considering this option in the future (within the next five years possibly). The main concerns with a cloud hosted NAV system from the user’s point of view (as noted in the survey) are bandwidth; uptime; data extraction for local reporting, performance in the rural setting, privacy and location of data. Therefore, analysts and consultants need to look further into these issues, and provide applicable solutions so that some of the companies without cloud based systems can consider the cloud option in the future.

**Limitations and Lessons from the Project**

With ERP implementations, there may be many other unpredictable variables (Marginson, 2008). Every implementation decision can make an impact in terms of project costs and the eventual real world use of the ERP systems. IT professionals can also learn a lot from the articles in the literature. There are continued motives to do more research in this field as some industries (for example higher education in New Zealand) are experiencing significant changes and adoption of new information systems (Hong, Katerattanakul, & Lee, 2014). However, it is important for every industry to analyse and satisfy the gaps between the ERP systems and their particular user requirements. This research can help organisations and specialists, including vendors, to help identify solutions to any potential disappointments of ERP system.

Overall, the survey results show that NAV has significant benefits, some of which depend on in-house maturity and greater resources in the future, as well as anticipated upgrades and new features. Some of the local users are in the early stages of the implementation and have not been able to answer all questions. Further analysis in the future will help these organisations and the ERP experts, achieve more, provide better service to the stakeholders, and take better advantage of ERP systems (Neilson, 2009). This paper determines that users need to focus more on ERP adoption and end user issues to make ERP systems more friendly and effective.

## Recommendations

Experts in ERP and change management share their advice on how to choose and implement an ERP system for a better return on the financial and time investment. Implementing an ERP system is an expensive plan, not just outsourcing implementation and maintenance, but in terms of internal resources and time. ERP users are satisfied if the vendors provide good results to the organisation. Implementing an ERP system helps in manufacturing business processes, for example, by increasing profits (long-term or even short-term). ERP systems allow organisations to manage financials, stocking, purchasing, selling, accounting, supply chain and so forth. In order to help organisations, increase the chances of a successful ERP implementation, an understanding of change management in itself is also important. Organisational change management is essential to the success of your ERP implementation, according to Matt Thompson, VP of Professional Services, Estes Group (an ERP innovation and solution provider) (Schiff, 2014). Run of the mill ERP ventures encourage enormous change in organisations that can change of everyday sets of job responsibilities or wiping out certain job tasks altogether. It is important for companies to address these changes during training workshops. Otherwise, there can be an unfriendly response to the ERP system during adoption.

**Suggestions for future research**

It may be interesting for future researchers to look at ERP systems just from the change management perspective. For any business or an organisation, upper management support is necessary. Research by Daniele Fresca as mentioned by Schiff (2014), shows that organisations that tend to fight the most against ERPs are those where the implementation project does not get enough upper level management support. Furthermore, Schiff (2014), business and technology writer and contributor to CIO.com, says that "personnel at the lower level tend not to be dedicated and involved with the implementation plan without senior level involvement". The management or the managers of the organisation must be aware of the issues related to ERP implementation process (but not necessarily about all of the configuration details).

Managers of the organisation must know their requirements well before choosing their ERP vendors. “The managers should focus on particular business processes and system requirements” (Schiff, 2014). A more accurate identification of functional requirements for ERP system implementation may help choose the right vendor. "Few things disrupt project funds and timelines as the 'assumptive' or absent requirements” (Schiff, 2014). Accordingly make sure to involve with end users, IT and senior management. Mostly, people select an ERP system based on reasons such as cost, present technology buzz or the system that is the flashiest.

"In any case, without a solid match, organisations are left with costly customization and integrated systems" (Schiff, 2014). According to Buyya, Broberg and Goscinski (2010), companies need to "find an ERP system that is industry specific, with devices and components intended to do the business necessities. The ROI and long term advantages of a well-fitting system are broad.” This project identified NAV as being relevant to the local industries; however, other future researchers may want to look at other ERP products due to their own geographical and industrial settings.

Furthermore, ERP systems should not forget mobile users. 'Bring your own device' or BYOD, is one of the major trends in the current business and computing environment (Erturk & Fail, 2013). Bring-your-own-device (BYOD) usage is increasing in businesses, and providing access to ERP systems requires more frequent use of virtualisation and remote desktop protocols (Buyya et al., 2010). Companies should pick an ERP system that allows clients to use cell phones and tablets, and yet in the meantime will ensure that sensitive data is secure. An interesting future research topic may be about the effective mobile use of ERP software.

**Suggestions for ERP Implementations**

"Absence of cooperation and contribution from key partners in the evaluation stage can prompt poor acknowledgment and client selection. Keep in mind that delays running the assessment can postpone the go-live date and an ideal opportunity to advantage” (Schiff, 2014). Tiffani Murray, a human resource technology consultant states, "another thing many organisations miss during the implementation phase of an ERP system is reporting and performance measurement” (Schiff, 2014). It is possible that this is conceivable by means of the current, prebuilt reports in the system. Otherwise the company needs to pay additional money to get customization that will drive the business, employing and resourcing. An ERP system that does not work with the current legacy and/or basic office systems is not an answer but rather another costly bit of unused or unusable software design, according to Buyya et al. (2010).

Companies that have not implemented ERP solutions yet should talk to other companies and "ask them what went right, what went wrong and what they might have done differently. If a vendor can't provide at least three verifiable, happy customers, they may not have the experience you need” (Schlichter & Kraemmergaard, 2010).

One place for these conversations is the local user's group (e.g. the Hawke's Bay NAV users' group). Therefore, it is important for local companies to continue these meetings and groups in the future, for example, by sponsoring them and sending their employees there.

Companies need to appoint an internal ERP champion and surround him/her with competent people. The principal, Morris Tabush at the TabushGroup says that "Try not to rely on the vendor designated project manager; have somebody on your staff for this" (Schiff, 2014). This individual will be in charge of gathering all the end user requirements, taking in the new system all around, working with the vendor on information processing, planning, preparing, and contacting all the other representatives (Rong, 2013). The ERP champions may conduct online surveys inside their companies in the future that can help them document some of the problems ahead of time.

Companies should provide the necessary time and resources for training on the ERP system. Learning a new way of doing work will demand time and commitment from all employees. According to Joel Schneider, Liberty Technology Advisors, an IT consulting firm with great experience in ERP implementations, the development team must find ways to decrease the load on employees. For example, well informed workers in each department can train the others (Motahari-Nezhad, Stephenson, & Singhal, 2009).

Finally, it is important to understand the training needs and create suitable training programs. Another idea for a future research project may be the qualitative analysis of a company's ERP system training documentation and training techniques. Future researchers may try to compare or evaluate the effectiveness of different training approaches and types of training materials, in terms of end user satisfaction and the achievement of business goals.